\documentclass[review]{elsarticle}
\usepackage{lineno,hyperref}
\usepackage{caption}
\usepackage{subcaption}
\usepackage{amsmath}
\usepackage[ruled]{algorithm2e} 

\modulolinenumbers[5]

\usepackage{graphicx}
\usepackage{multirow}
\usepackage{lineno,hyperref}

\let\linenumbers\nolinenumbers\nolinenumbers
\begin{document}

\begin{frontmatter}

\title{Representation Extraction and Deep Neural Recommendation for Collaborative Filtering}


\author[secondaddress]{Arash Khoeini}
\ead{akhoeini@sfu.ca}
\author[mymainaddress]{Saman Haratizadeh \corref{cor1}}
\ead{haratizadeh@ut.ac.ir}
\cortext[cor1]{Corresponding author}
\author[secondaddress]{Ehsan Hoseinzade}
\ead{ehsan\_hoseinzade@sfu.ca}

\address[mymainaddress]{Faculty of New Sciences and Technologies, University of Tehran, North Kargar Street, 1439957131 Tehran, Iran}
\address[secondaddress]{School of Computing Science, Simon Fraser University, 8888 University Drive, Burnaby, Canada}

\begin{abstract}
Many Deep Learning approaches solve complicated classification and regression problems by hierarchically constructing complex features from the raw input data. Although a few works have investigated the application of deep neural networks in recommendation domain, they mostly extract entity features by exploiting unstructured auxiliary data such as visual and textual information, and when it comes to using user-item rating matrix, feature extraction is done by using matrix factorization. As matrix factorization has some limitations, some works have been done to replace it with deep neural network. but these works either need to exploit unstructured data such item's reviews or images, or are specially designed to use implicit data and don't take user-item rating matrix into account.
In this paper, we investigate the usage of novel representation learning algorithms to extract users and items representations from rating matrix, and offer a deep neural network for Collaborative Filtering. Our proposed approach is a modular algorithm consisted of two main phases: REpresentation eXtraction and a deep neural NETwork (RexNet). Using two joint and parallel neural networks in RexNet enables it to extract a hierarchy of features for each entity in order to predict the degree of interest of users to items. The resulted predictions are then used for the final recommendation. Unlike other deep learning recommendation approaches, RexNet is not dependent to unstructured auxiliary data such as visual and textual information, instead, it uses only the user-item rate matrix as its input. We evaluated RexNet in an extensive set of experiments against state of the art recommendation methods. The results show that RexNet significantly outperforms the baseline algorithms in a variety of data sets with different degrees of density.

\end{abstract}


\end{frontmatter}

\linenumbers

\section{Introduction}

The massive amount of available products in e-commerce websites had made Recommendation System a crucial need for those businesses. Collaborative Filtering is the dominant approach for recommendation when the sole available data is the user-item interactions matrix. The core idea of CF methods is to estimate the favorite items for a specific target user, based on previous implicit or explicit evaluations of the users about items of the system. 

A recent approach in recommendation systems is deep Learning. Deep learning (\cite{lecun2015deep}) methods have been extremely successful in different research areas like image processing (\cite{krizhevsky2012imagenet}), natural language processing (\cite{mikolov2010recurrent}), face detection (\cite{schroff2015facenet}) and stock market prediction (\cite{hoseinzade2019cnnpred}, \cite{hoseinzade2019u}). 
Inspired by these  successes of deep learning algorithms in mentioned domains, some researchers have tried to introduce novel deep learning approaches for recommendation systems. These methods use deep models such as deep neural networks to map the available data about users and items to a prediction about unknown user-item relations. For example, some approaches use reviews written by users about item to capture representations for those entities through deep models  (\cite{zheng2017joint} , \cite{paradarami2017hybrid}). Some other approaches integrate different data sources e.g. images, texts, etc to capture representations (\cite{nedelec2017specializing} , \cite{zhang2016collaborative}) to estimate user preferences.
Although there have been various successful works among these attempts, they heavily rely on using unstructured content data, such as reviews or images, mainly because the deep models have already been designed to handle such sort of data. This makes these approaches inapplicable when the only available source of information is user-item rating matrix\cite{bennett2007netflix}. 

Deep Neural Networks are efficient tools for automatic feature extraction probably due to their ability of extracting features hierarchically \cite{krizhevsky2012imagenet}. In this hierarchical feature extraction approach, each layer aggregates the features that are extracted by its previous layers into a higher level features. These neural networks usually have a tower architecture, where there is fewer number of neurons in top layers.

In this paper we introduce Deep Joint Feature eXtraction (RexNet), a novel recommendation approach which is composed of three main training steps. First, it transform the user-item matrix into sentence-like sequences. Each sequence comprises all the items that a user has shown interest to. In the second step, RexNet uses the sequences of last step to extract representations for items and users using a representation learning method, such as Skip-gram (\cite{mikolov2013distributed}) or GloVe (\cite{pennington2014glove}). Although Skip-gram  has already been adopted for content-based representation learning in recommender systems, here we suggest to apply GloVe for entity representation learning from user-item data. For the third step, a deep feed forward neural network structure is introduced which is composed of two parallel deep neural networks that are connected through their output neuron. Each of these neural networks are responsible for implicitly extracting features for either users or items. They receive the captured representations of a user-item pair from previous step, and are trained to jointly predict the unknown relation between users and items through their common output neuron. Based on these predictions the system can sort the items for any target user and recommend the top rated items to that user.

The contributions of this work can be summarized as below:

\begin{itemize}
    \item suggesting a new model-based framework for recommendation in collaborative filtering domain that outperforms other state of the art approaches.
	\item Suggesting a new approach for using GloVe and SkipGram, to extract vector representations for users and items when the sole available information is the user-item  relation matrix.
	\item Proposing a special deep neural network architecture that uses the extracted vectors for each user/item pair, to implicitly extract a hierarchy of features for them and then predict the unknown interest of the user about the item.
\end{itemize}

\section{Related Work}
Collaborative Filtering is the most successful and applied approach in Recommendation Systems. Collaborative Filtering methods use users previous interactions to find similar users or items to recommend a target item to the target user, and there are various approaches that can be used to reach this goal. For instance, matrix factorization methods (\cite{mnih2008probabilistic}, \cite{koren2009matrix}, \cite{xiao2018coupled}, \cite{jamali2010matrix}, \cite{xu2018novel}, \cite{meng2018search}) factorize user-item interaction matrix to estimate the unknown user-item interactions, or graph-based approaches (\cite{shams2018item}, \cite{shams2018reliable}, \cite{shams2017graph}, \cite{musto2017introducing}, \cite{najafabadi2019impact}) use entities relation networks to find similar users and items.  

One of first attempts on using neural networks in recommendation area was done by Geoffery Hinton et. al. in 2007 (\cite{Salakhutdinov:2007:RBM:1273496.1273596}). In this work authors used Restricted Boltzmann Machine (RBM) to model users rating of movies. They proposed an efficient inference procedure that could outperform SVD, which was the dominant method of Collaborative Filtering. A more recent usage of RBM in improving recommendation systems is \cite{hazrati2019entity}, where authors used RBM to learn representation for users and preferences. Another attempt that used deep neural network in recommendation area was done by Sander Deileman et. al.\cite{van2013deep} where they trained a Convolutional Neural Network(CNN) to recommend musics to users. They trained the CNN to predict the music latent factors from its audio signal. This attempt is one of the examples of training a deep model by content information to improve recommendations. Another example is \cite{zheng2017joint} in which authors used items reviews to train to parallel CNN. One of these networks is in charge for modeling latent factors of users and another one for items. These two networks were connected in a shared fully connected layer and in the shared layer a Factorization Machine were used to predict users unknown rankings. Recurrent Neural Networks are also used in the domain of recommendation systems. For example, \cite{hidasi2015session} applies RNNs to session-based recommender systems, or \cite{li2019characterizing} uses LSTM networks (\cite{hochreiter1997long}) for the the task of predicting downloads in academic search.

As deep neural networks are very powerful in extracting features from unstructured data, some recent methods have designed  modular models which combine different source of information such as text, image and audio. In one of these works \cite{nedelec2017specializing} authors propose an architecture which ends in a fusion step. This fusion step helps the model to keep a modular architecture where each module is trained to use a different kind of content information, such as images or texts. 
The other example is \cite{zhang2016collaborative} where authors use different auxiliary information to solve the sparsity problem, which is one main challenge in collaborative filtering. They adopt a heterogeneous network embedding method, termed as TransR, to extract items' structural representations by considering the heterogeneity of both nodes and relationships. They apply stacked denoising auto-encoders and stacked convolutional auto-encoders, which are two types of deep learning based embedding techniques, to extract items' textual representations and visual representations, respectively.

Although the mentioned attempts have been successful in using auxiliary information, they are inapplicable when the only available source of information is user-item rating matrix. Some other works has been done to use neural networks in these situations, when all we know about users and item is just a rating matrix. A notable approach is NeuMF (\cite{he2017neural}) where authors propose a neural network to generalize matrix factorization on implicit data. An example work of using neural network on rating data is \cite{Sedhain:2015:AAM:2740908.2742726}, were authors propose AutoRec, which is a novel autoencoder framework for collaborative filtering. AutoRec can be user-based or item-based. In each situation it uses an Auto Encoder to learn compact representations of users or items. AutoRect then uses these compact representations to estimate user or item full vector of rating estimations. 


Another example of learning representation to do recommendation is \cite{barkan2016item2vec} where items are mapped to a dense vector space using Skip-gram with Negative Sampling and then a recommendation is done by a nearest-neighbor approach.

Skip-gram with Negative Sampling and Glove are two main Representation Learning approaches in Natural Language Processing. As we will use these methods in our work, we will explain both methods briefly in the following sub sections.

\subsection{SkipGram with Negative Sampling}

	\begin{figure}[ht]
	\center
    \includegraphics[width=0.3\textwidth]{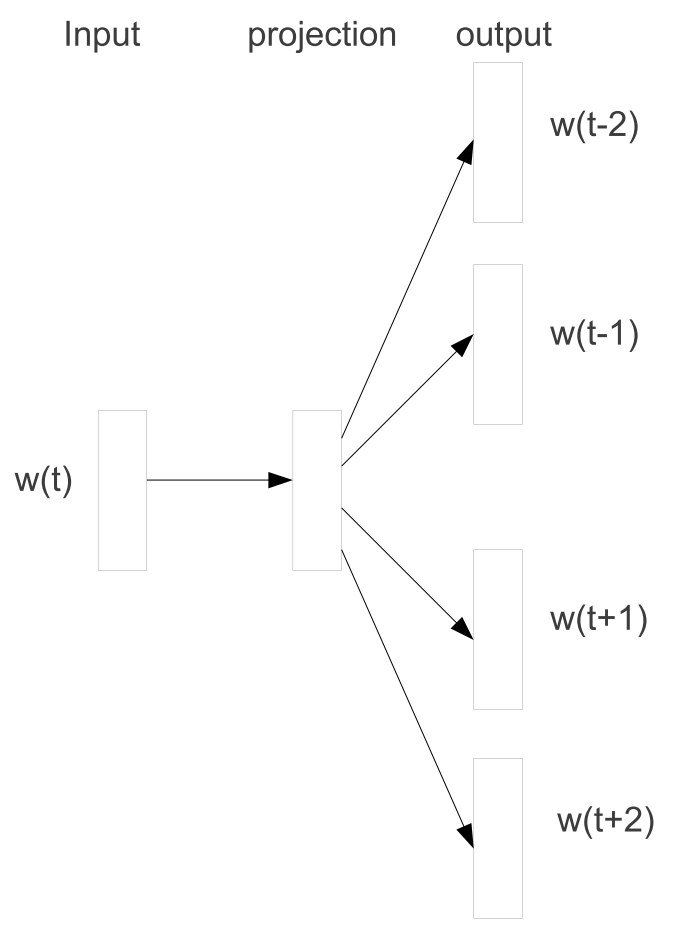}
    \caption{The Skip-gram model architecture. The training objective is to learn word vector representations that are good at predicting the nearby words \cite{mikolov2013distributed} }
    \label{chart2}
  \end{figure}

Skip-gram is a Word2Vec tool which was introduced in \cite{mikolov2013distributed}. Skip-gram architecture is a neural network with a hidden layer which tries to maximize function \ref{skipgram}.

	\begin{equation}  \label{skipgram}
		\frac{1}{T} \sum^T_{t=1} \sum_{-c \le j \le c, j \ne 0} log p(w_{t+j} | w_j)
	\end{equation}

	where c is the size of training context. Skip-gram defines $p(w_{t+j} | w_j)$ using the softmax function: 

	\begin{equation}  \label{softmax}
		p(w_O | w_I) = \frac {exp(v'^T_{w_O} v_{w_I})}  {\sum^W_{w=1} exp(v'^T_w  v_{w_I})}
	\end{equation}

	where $v_w$ and $v'_w$ are the input and output vector representations of w and W is the the size of vocabulary. Since this approach is impractical for very large vocabularies, Negative Sampling is proposed in original paper. Skip-gram with negative sampling has the objective function \ref{sg_ns} which is used to replace every $log P(w_O|w_I)$ term.

	\begin{equation} \label{sg_ns}
		log \sigma (v'^T_{w_O} v_{W_I}) + \sum^k_{i=1} E_{w_i \sim P_n(w)} [log \sigma (-v'^T_{w_i} v_{w_I})]
	\end{equation}

	Where k is number of negative samples, $\sigma$ is the Sigmoid function and $p(w)$ is a probability distribution for negative samples. 

	Skip-gram slides a fixed-size window  over corpus sentences and tries to maximize equation \ref{skipgram} on each window location. After training,  the matrix $W$ will contain captured representation for words in vocabulary based on their co-occurrences.  Using a fixed-size window size seems reasonable in Natural Language Processing tasks where the aim is to capture word similarities based on appearance of words in a word sequence such as a sentence or a paragraph. However it might limit the model from utilizing some co-occurrences in Collaborative Filtering systems in which usually such a sequential structure does not exist. 

	\subsection{GloVe}

	Although GloVe captures word representations for words as Skip-gram, it uses a different approach to achieve this goal. GloVe first scans the whole corpus of sentences and builds a matrix X, where $X_{ij}$ is the number of times word $i$ has been seen in the context of word $j$. Context is defined in a fixed-size window of words here and its size is a hyper parameter of the algorithm. After building matrix X, GloVe learns word vectors by minimizing the following cost function: 

\begin{equation}\label{glove_cost}
    J = \sum^V_{i,j=1} f(X_{ij}) (w^T_i \tilde{w}_j + b_i + \tilde{b}_j - \log X_{ij})^2
\end{equation}

where $f(x_{ij})$ is defined as follows: 

\begin{equation}
    f(x) =
    \left\{
        \begin{array}{ll}
            (x/x_{max})^{\alpha}  & \mbox{if } x < x_{max} \\
            1 & \mbox{if } otherwise
        \end{array}
    \right.
\end{equation}
	
Although GloVe uses fixed-size window to build matrix $X$, it gives the flexibility to change the definition of $X$ as long as it satisfies some constrains ( \cite{pennington2014glove} ). In our method, we will use GloVe to capture base representations for users and items, which will be fed to our deep neural network. 

\section{RexNet}

In the big picture, RexNet first captures dense representations of users and items. Then the captured representations are used to train a neural network for user-item relevance prediction. 

To capture dense representations, we can use GloVe or Skip-gram with Negative Sampling (SGNS). As we mentioned before, these methods are used in Natural Language Processing tasks, and they use massive sets of sentences as input. So in order to use any of these methods, first we need to build a corpus. One way of building that is defining sentences from user profiles. For example, if a user has rated twelve items, this user will add a sentence of size twelve to our corpus, where each word represents an item. The problem with this method is that the sentence structure will not distinguish low-rated items form high-rated ones. To address this problem, we allow each item appear in the sentence several times. The number of appearance is equal to the rate given to that item by the corresponding user. 

\begin{figure}[ht]
    \centering
  \begin{subfigure}[b]{0.4\textwidth}
    \includegraphics[width=\textwidth]{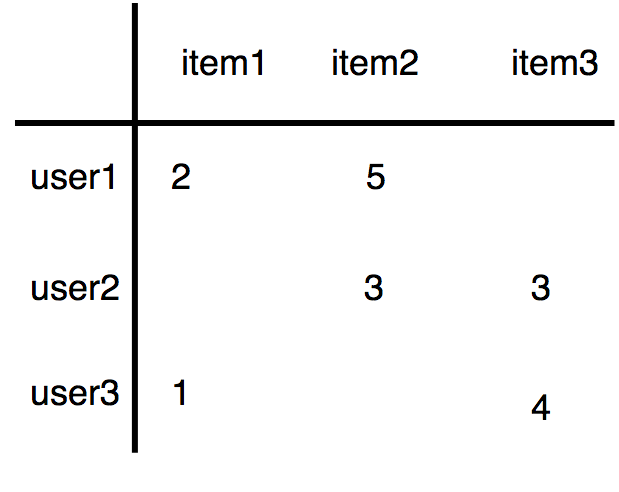}
    \caption{An example of user-item rating matrix}
    \label{useritemmatrix}
  \end{subfigure}
  \centering
  \hfill
  \begin{subfigure}[b]{0.5\textwidth}
    \includegraphics[width=\textwidth]{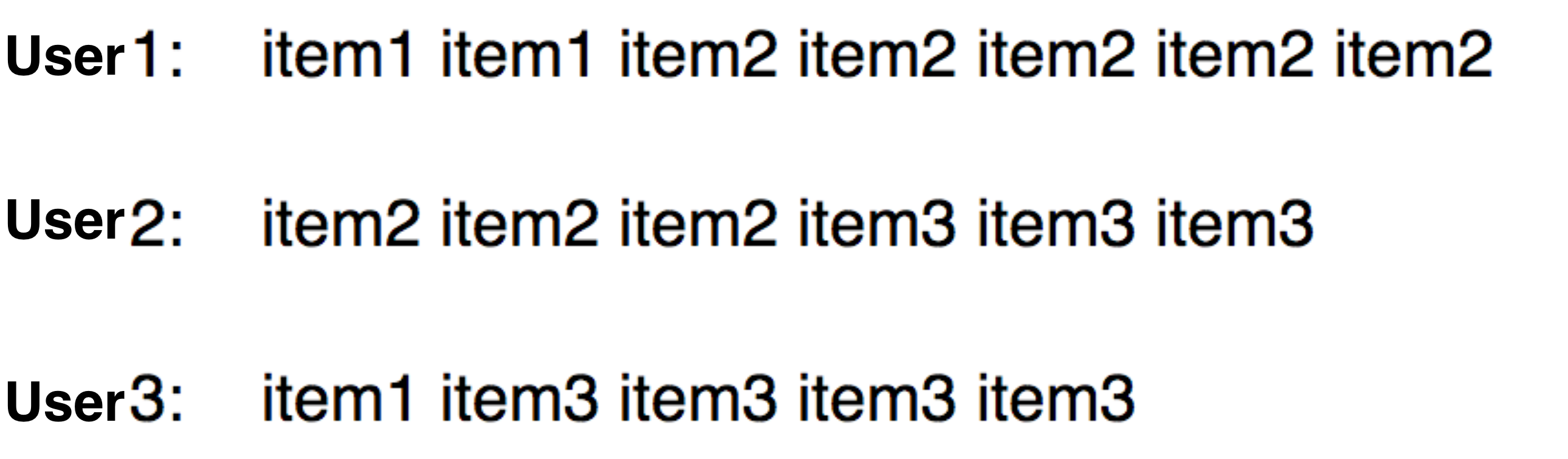}
    \caption{Extracted sentences from user-item rating matrix}
    \label{sentences}
  \end{subfigure}
\end{figure}

Figure \ref{useritemmatrix} shows an example of a user-item rating matrix and figure \ref{sentences} shows the extracted sentences of that matrix. 

Another concern about the transforming the profiles to sentences is that usually there is no sequence among items in a user profile while a sentence is naturally a sequence of words. To avoid any biases that this may cause, we use a shuffled version of the resulted sentences. 

After building a corpus of items, we can use this corpus to train GloVe or SGNS and capture dense representations for users. Since the prediction model needs users' representations as well, we need a way to derive user representations. One interesting characteristics of representations learned by GloVe or SNGS, is that they allow vector-oriented reasoning based on the offset between entities. For example when we are dealing with word entities, the male/female relationship is automatically learned, and with the induced vector representations, "King - Man + Woman" results in a vector very close to "Queen" \cite{mikolov2013linguistic}. Using this view, we argue that user item relationship is automatically learned when we use these techniques in Collaborative Filtering, and a user's taste can be define by all the items they have rated. So a user's representation can be defined as equation \ref{uservector}.

\begin{equation} \label{uservector}
    u_i = \sum_{k \in M_i} (r_{i,k} - \lambda_{i}) v_k ,
\end{equation}
where  $M_i$ is the set of items that user $i$ has rated. $r_{i,k}$ is the rate user $i$ has given to item $k$ and $\lambda_{i}$ is the average of ratings given by user $i$.

As the third module, we introduce a deep and parallel feed forward neural network to extract hierarchical features of items and users. Using neural networks for hierarchical feature extraction have shown massive success in different areas of deep learning. These neural networks usually have a tower structure, where bottom layers are the widest and each successive layer has smaller number of neurons. The initial wider layers of the neural network extract more general and basic features while the higher layers with small number of hidden units extract more abstract features \cite{he2016deep}.

\begin{figure}[ht]\caption{Parallel Feed Forward Neural Network: the upper one is the user network and lower one is the item network. These two networks are connected in the shared layer.}
    \includegraphics[width=\textwidth]{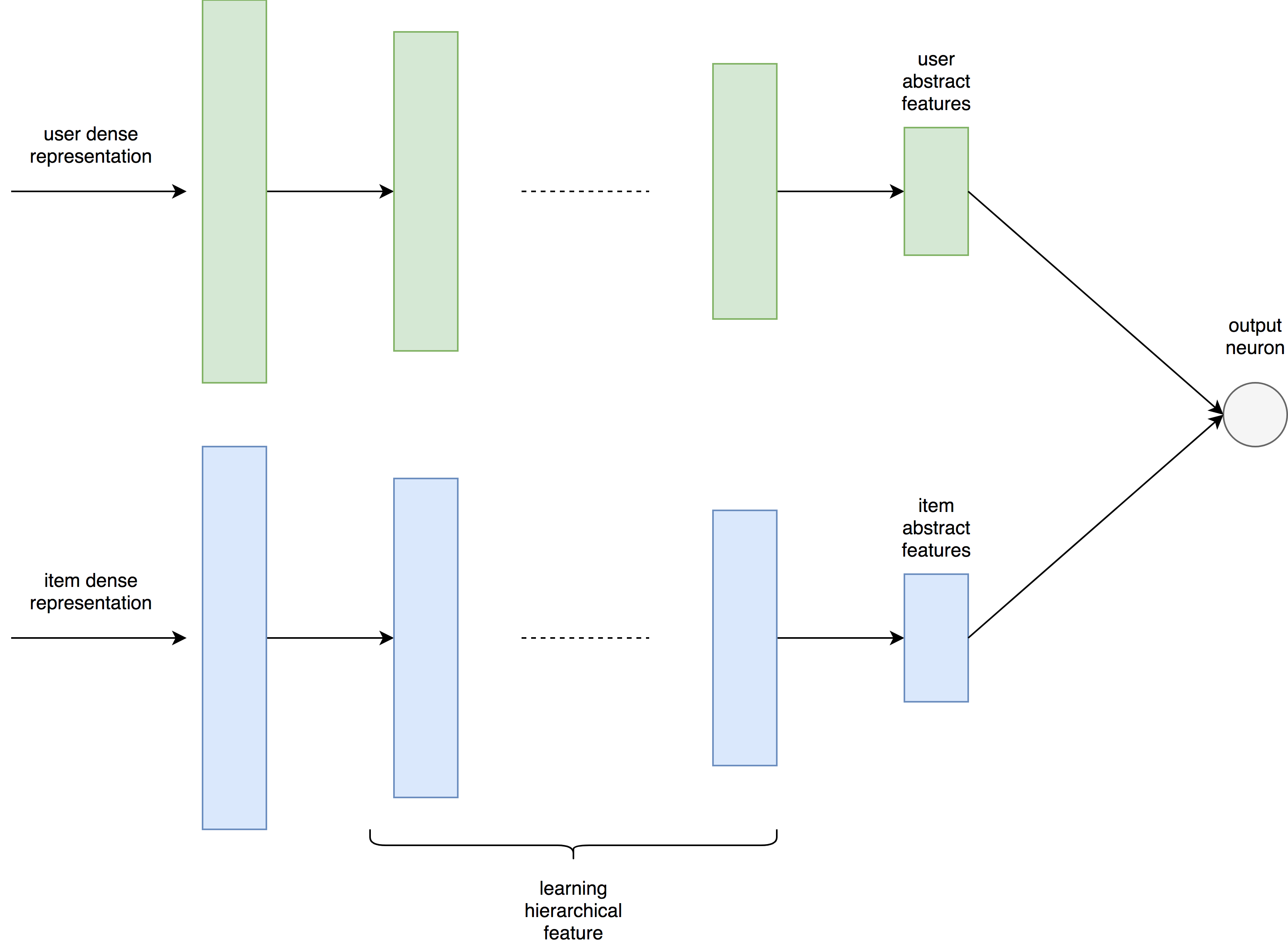}\label{ParallelNN}
\end{figure} 

The architecture of the proposed neural network is presented in figure \ref{ParallelNN}. As we can see, this neural network consists of two parallel deep multi-layer perceptrons which are connected in one shared output neuron. The aim of choosing this architecture is to give our model the opportunity to extract features for users and items separately. As we can see in figure \ref{ParallelNN}, the upper module of the neural network receives the item's base representations in its input and extract items' features in several layers. The user's module does the same thing for users. The number of layers and size of neurons in each layer are hyper parameters of the model and depend on the properties of the available training data. Being connected in the output neuron, enables the model to back-propagate the same final error through the layers of both modules. Using this approach of learning forces both modules to extract features in a way that the final interaction yield the user preference for the item. 

To train this neural network, we feed it with a user and an item base representations that are captured as we explained before. The target value which the neural network should learn to estimate is the preference level of the user about the item. We have defined the preference as the rate given by the user to that item minus the average of ratings given by the user. For example if a user with average rating of 3.5 has rated an item 4, the preference will be 0.5 and our neural network will learn to estimate it. Output neuron uses linear combination of final high level features of users and items to estimate the preference. If we consider $a^u_n$ the output vector of the last hidden layer in user's neural network (user's abstract feature vector), and $a^i_n$ the output vector of the last hidden layer in item's neural network (item's abstract feature vector), the final output will be calculated as function \ref{lastlayer_fucntion}

\begin{equation} \label{lastlayer_fucntion}
	f(u_i, v_j) = V \left[ \begin{array}{c}a^u_n \\ a^i_n \end{array} \right] + b_{n+1}
\end{equation}
where $V$ is the weight vector of the output neuron and $a^i_n$ and $a^u_n$ are defined as functions \ref{user_abstract_layer} and \ref{item_abstract_layer} respectively. In these functions, $u_j$ denotes the captured representation for user $j$ and $v_k$ denotes the captured representation for item $k$.

\begin{equation} \label{user_abstract_layer}
	a^n_u = ReLU(W^u_n ReLU(W^u_{n-1}(...ReLU(W^u_1 u_j + b^u_1)...)+b^u_{n-1})+b^u_n)
\end{equation} 

\begin{equation} \label{item_abstract_layer}
	a^n_i = ReLU(W^i_n ReLU(W^i_{n-1}(...ReLU(W^i_1 v_k + b^i_1)...)+b^i_{n-1})+b^i_n)
\end{equation} 

Where $W^u_j$ and $W^i_j$ are weights of jth layer of user's and item's neural network respectively. Biases of these layers are also indicated by $b^u_j$ and $b^i_j$.

\begin{figure}[ht]\caption{RexNet} \label{RexNet_figure}
    \includegraphics[width=\textwidth]{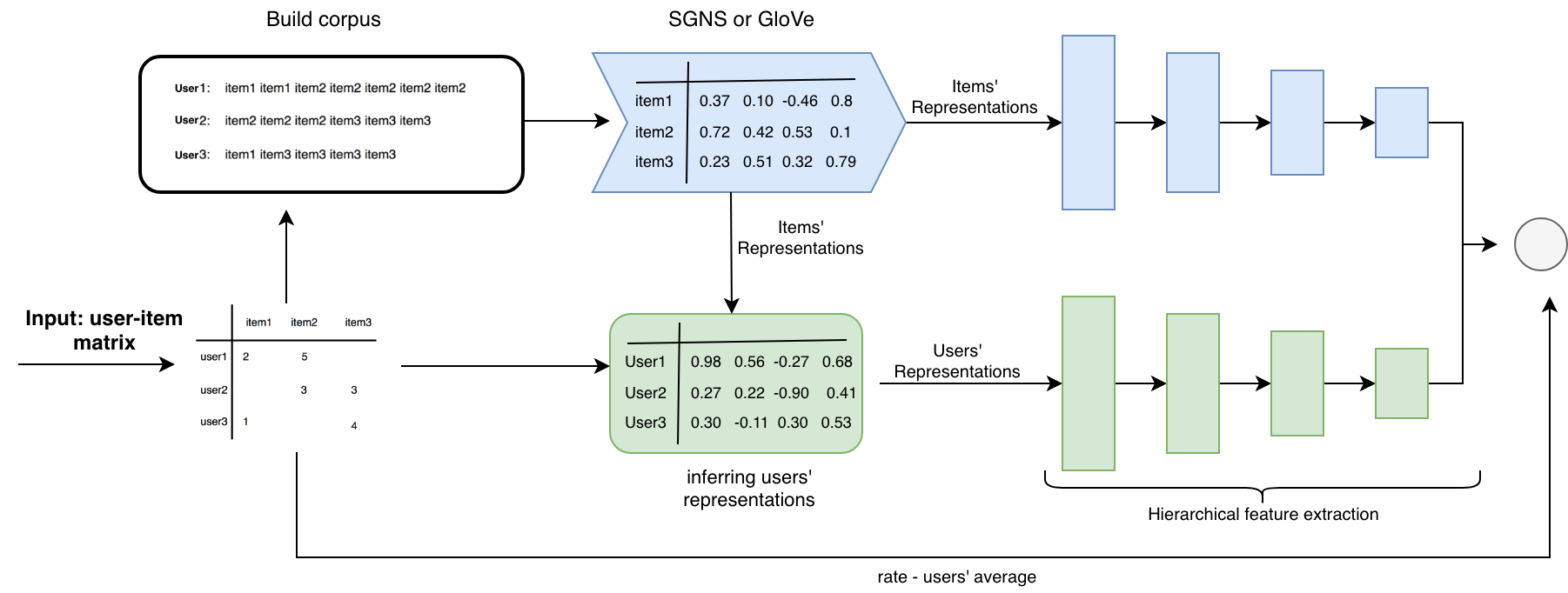}\label{ParallelNN}
\end{figure}

Figure \ref{RexNet_figure} illustrates whole RexNet structure. As we can see, the algorithm's input is only user-item rating matrix. As we explained earlier, first this matrix is turned to a sentence-like sequences and then a deep representation learning method is used to capture item representations. After that, these representations and user-item matrix are used to capture users' representations based on equation \ref{uservector}, and finally all these representations are used to feed and train a deep joint neural network. This procedure is also explained more formally by pseudo-codes. Pseudo-code \ref{algorithm_module1}, \ref{algorithm_module2} and \ref{algorithm_module3} show the procedure of three modules respectively, and algorithm \ref{algorithm_main} is the main function of RexNet that uses all other modules.

\begin{algorithm}
\caption{RexNet main}\label{algorithm_main}
\KwIn { user-item rating matrix}
\KwOut {trained joint neural network}
\BlankLine
\tcp{build corpus}
\textit{sentences} $\gets$ build-corpus(\textit{user-item-ratings}) 
\\
$\textit{user-averages} \gets \text{calculate-average(\textit{user-item-matrix})}$
\\
\tcp{capture users' and item's representations}
\textit{item-representations} , \textit{user-representations} $\gets$ capture-representations(\textit{sentences}, \textit{user-averages})
\\
\tcp{train deep joint neural network}
\textit{neural-model} $\gets$ \text{train-deep-neural-network}(\textit{user-item-rating}, \textit{item-representations}, \textit{user-representations}, \textit{user-averages} )
\end{algorithm}

\begin{algorithm}
\caption{RexNet Module 1: build corpus}\label{algorithm_module1}
\KwIn { user-item-rating matrix}
\KwOut {corpus}
\BlankLine
\SetKwFunction{FMain}{build-corpus}
\SetKwProg{Fn}{Function}{:}{}
  \Fn{\FMain{\textit{user-item-rating}}}{
        $\textit{sentences} \gets initialize-list(size= \text{number-of-users} )$
        \\
        \ForEach{rate[user,item] $\in$ user-item-rating}
        {
        $sentences[user].append(\text{repeat item \textit{rate} times}) $
        }

        \KwRet sentences
  }
\end{algorithm}
\begin{algorithm}
    \caption{RexNet Module 2: capture representations}\label{algorithm_module2}
    \KwIn { sentences, user-averages}
    \KwOut {users' and items' representations}
    \BlankLine
    \SetKwFunction{FMain}{capture-representations}
    \SetKwProg{Fn}{Function}{:}{}
      \Fn{\FMain{\textit{user-item-rating}, user-averages}}{
            \tcp{using SGNS or GloVe as representation learning algorithm}
            $\textit{item-representations} \gets 
            \text{representations-learning(\textit{sentences})}$

            \tcp{infer user representations using equation \ref{uservector}}
            \ForEach{user $\in$ users}{
            $ \textit{user-rated-movies , rates} \gets \text{get-movies-and-rates(\textit{user})} $ 
            \\
            $ \textit{user-vector} = (\textit{rates} - \textit{user-averages}[\textit{user}]) * \textit{item-representations}[\textit{user-rated-movies}] $
            $\textit{user-representations}.append(\textit{user-vector)}$
            }
    \KwRet \text{item-representations, user-representations}}
\end{algorithm}
\begin{algorithm}
\caption{RexNet Module 3: train deep joint neural network}\label{algorithm_module3}
\KwIn { user's and items' representations, user-averages}
\KwOut {deep joint model}
\BlankLine
\SetKwFunction{FMain}{train-deep-neural-network}
\SetKwProg{Fn}{Function}{:}{}
  \Fn{\FMain{\textit{user-item-rating}, \textit{item-representations}, \textit{user-representations}, \textit{user-averages}}}{
        
        $\textit{neural-network-data} \gets \text{initialize-list()}$
        \\
        \ForEach{\textit{rate[user,item]} $\in$ \textit{user-item-rating}}{
        
        \textit{neural-network-data}.append((\textit{user-representations}[\textit{user}], \textit{item-representations}[\textit{item}], \textit{user-item-matrix}[\textit{user},\textit{item}] - \textit{user-averages}[\textit{user}]))
        
        }
        
        \tcp{train joint neural network}
        $ \textit{neural-model} \gets \text{build-and-train-deep-neural-network(\textit{neural-network-data})}$

        \KwRet \text{trained-deep-model}
  }

\end{algorithm}

\section{Experiments}
We compared the performance of RexNet against different methods of Collaborative Filtering on two datasets with different degrees of sparsity: MovieLens 100k and MovieLens 1M datasets. We used the rating data of these datasets and ignored timestamps. MovieLens-100K contains 100 thousands ratings  of 943 users with at least 20 rated movies for each user and there are 1682 distinct movies in it. MoveiLens-1M contains 1 Million ratings by 6040 users with at least 20 rated movies for each user. There are 3952 distinct movies in MoveiLens-1M. 

For each user, a fixed number (UPL) of movies that he/she has rated are selected for training and the remaining ratings are left for the testing process. A user’s profile size is the number of items in his/her train set. Different sizes of user profiles are used to compare our method with baseline methods.  We used UPL sizes of 10, 20 and 50 in our experiments.

\subsection{Parameter Setting}
For the representation learning, we fixed the representation size to 100 and the window size to 25 and did not explore other sizes as our main focus was in the performance of the neural network module. We chose to use four hidden layers for each neural networks with 30, 20, 10, and 5 neurons for the first through the last layer. All hidden layers use ReLU activation function and the output neuron has no activation function.

\subsection{Evaluation Metric}

We have used NDCG for evaluating the performance of the recommendation algorithms in our experiments. NDCG is a metric for measuring the performance of rank-based recommender systems, which is used to evaluate top-n recommendations in the recommendation lists. The more the top-n recommendations are similar to real top-n items ranked by the user, the NDCG gets closer to 1. NDCG for top-n items are calculated 
as below:

\begin{equation} \label{NDCG}
    NDCG_u = \frac{1}{\beta_u} \Sigma^{top-n}_{i=1} \frac{2^{r^u_i} - 1} {log^i_2 +1}
\end{equation},

where $r^u_i$ is the rate of the $i^{th}$ ranked item and $\beta$ is a normalizing factor that makes sure that NDCG is between 0 and 1.

\section{Baseline Algorithms}

In our experiments, RexNet is compared with the following model-based Collaborative Filtering algorithms:
\begin{itemize}
     \item AutoRec \cite{Sedhain:2015:AAM:2740908.2742726} that is a neural-network-based algorithm which uses an Auto Encoder to estimate user ratings.
    \item ListRank \cite{shi2010list}  which is a ranking-oriented Collaborative Filtering approach.

    \item CofiRank \cite{weimer2008cofi}, that uses a Matrix Factorization Collaborative Ranking approach.

    \item PushAtTop \cite{christakopoulou2015collaborative} which is a Matrix Factorization algorithm which weights pairwise comparisons according to their position in the total ranking of items for users. PushAtTop contains three different algorithms: push-inf, push-p, push-reverse.
\end{itemize} 
\subsection{Parameter Setting}
The input layer of RexNet has 100 neurons in both modules, which means embedding vectors of users and items also have 100 dimensions. We used three hidden layers for both user's and item's networks where the number of hidden neurons are 30, 20, 10, and 5 respectively. 

We also used a Dropout layer of 0.4 after both hidden layers of size 5.  The number of neurons in the the shared layer is 5 and we have a Dropout layer of 0.2 after it. Output layer contains just one neuron with no activation function.
Neurons in all other layers use Rectified Linear Unit (ReLU) as activation function.
Parameters of the baseline algorithms are set as follows:

\begin{itemize}

    \item In AutoRec, the number of hidden neurons is 500, and the epoch number is 10.

    \item In ListRank, dimension of latent features (d) is 10 and regularization parameter is set to 0.1.

    \item In CofiRank, publicly available framework is used and the best parameters reported from \cite{weimer2008cofi} is used.

    \item For three different algorithms of PushAtTop, All the optimal algorithms reported in \cite{christakopoulou2015collaborative} are
    used.
\end{itemize}

\subsection{Results} 
Tables \ref{result100k} and \ref{result1M} show the NDCG of RexNet and other state-of-art algorithms on MoveLens 100K and 1M respectively. In each dataset  we trained RexNet using the embeddings captured by GloVe and Skip-gram separately. As we can see in both tables, RexNet performs better than all the baseline methods in terms of NDCG.

\begin{table*}[ht] \caption{Performance comparison in terms of NDCG between RexNet and state of the art recommendation algorithms on MovieLens 100k dataset.}
    \label{result100k}
    \centering
    \resizebox{\textwidth}{!}{

\begin{tabular}{ |c||c|c|c|c|c|c|  }

    \hline
     &              \multicolumn{2}{c}{UPL = 10}             &      \multicolumn{2}{c}{UPL = 20}       & \multicolumn{2}{c|}{UPL = 50} \\
    \hline
    algorithm  & NDCG@5             & NDCG@10             & NDCG@5             & NDCG@10             & NDCG@5             & NDCG@10  \\
    \hline
    RexNet (using GloVe)     & \textbf{0.680 $\pm$ 0.011}& \textbf{0.697 $\pm$ 0.005}& \textbf{0.690 $\pm$ 0.004}& \textbf{0.700 $\pm$ 0.003}& \textbf{0.707 $\pm$ 0.004}& \textbf{0.709 $\pm$ 0.004}\\
    RexNet (using Skip-gram) &  0.673$ \pm$0.010 & 0.693$\pm$0.012 & 0.681$\pm$0.010 & 0.691 $\pm$ 0.005    & 0.692 $\pm$ 0.009  & 0.695 $\pm$ 0.007  \\
    AutoRec    & 0.614 $\pm$ 0.002  & 0.648 $\pm$ 0.002   & 0.644 $\pm$ 0.001  & 0.671 $\pm$ 0.002    & 0.682 $\pm$ 0.003  & 0.692 $\pm$ 0.001  \\
    ListRank   & 0.672 $\pm$ 0.005  & 0.693 $\pm$ 0.005   & 0.682 $\pm$ 0.006  & 0.691 $\pm$ 0.004    & 0.687 $\pm$ 0.008  & 0.684 $\pm$ 0.004  \\
    CofiRank   & 0.602 $\pm$ 0.020  & 0.631 $\pm$ 0.013   & 0.603 $\pm$ 0.014  & 0.620 $\pm$ 0.024    & 0.609 $\pm$ 0.010  & 0.616 $\pm$ 0.004  \\
    push-inf   & 0.611 $\pm$ 0.030  & 0.640 $\pm$ 0.023   & 0.629 $\pm$ 0.015  & 0.647 $\pm$ 0.012    & 0.658 $\pm$ 0.003  & 0.667 $\pm$ 0.002  \\
    push-reverse& 0.594 $\pm$ 0.043  & 0.623 $\pm$ 0.033  & 0.621 $\pm$ 0.018  & 0.640 $\pm$ 0.016    & 0.664 $\pm$ 0.012  & 0.668 $\pm$ 0.013  \\
    push-p     & 0.519 $\pm$ 0.021  & 0.561 $\pm$ 0.015   & 0.580 $\pm$ 0.034  & 0.602 $\pm$ 0.029    & 0.681 $\pm$ 0.012  & 0.679 $\pm$ 0.011  \\
    
    \hline
   \end{tabular}
    }
\end{table*}

\begin{table*}[ht] \caption{Performance comparison in terms of NDCG between RexNet and state of art recommendation algorithms on MovieLens 1M dataset.}
    \label{result1M}
    \centering
    \resizebox{\textwidth}{!}{

\begin{tabular}{ |c||c|c|c|c|c|c| }

    \hline
     &             \multicolumn{2}{c}{UPL = 10}             &     \multicolumn{2}{c}{UPL = 20}        &      \multicolumn{2}{c|}{UPL = 50}     \\
    \hline
    algorithm     & NDCG@5             & NDCG@10            & NDCG@5            & NDCG@10             & NDCG@5            & NDCG@10            \\
    \hline
    RexNet (using GloVe) & 
    0.734 $\pm$ 0.004 & 0.738 $\pm$ 0.004 & 0.742$\pm$ 0.005 & 0.743$\pm$ 0.003 & 0.747 $\pm$ 0.001 & 0.746$\pm$ 0.001  \\
    RexNet (using Skip-gram) & 
    \textbf{0.741 $\pm$ 0.002} & \textbf{0.744 $\pm$ 0.001} & \textbf{0.746$\pm$ 0.001} & \textbf{0.747$\pm$ 0.001} & \textbf{0.751$\pm$ 0.001} & \textbf{0.752 $\pm$ 0.006}   \\
    AutoRec       & 0.625 $\pm$ 0.015          &      0.634 $\pm$ 0.008        &      0.627 $\pm$ 0.008    &     0.637 $\pm$ 0.007    &     0.648 $\pm$ 0.009     &     0.659 $\pm$ 0.006  \\    
    ListRank & 0.647 $\pm$ 0.002 &  0.654 $\pm$ 0.002 & 0.683 $\pm$ 0.003 & 0.688 $\pm$ 0.003 &  \textbf{0.751 $\pm$ 0.002} &  0.751 $\pm$ 0.002  \\
    CofiRank & 0.685 $\pm$ 0.002 &  0.684 $\pm$ 0.001 & 0.676 $\pm$ 0.008 & 0.685 $\pm$ 0.024 &  0.641 $\pm$ 0.005 &  0.644 $\pm$ 0.003  \\
    push-inf      & 0.690 $\pm$ 0.008          &      0.699 $\pm$ 0.002        &      0.691 $\pm$ 0.017    &     0.697 $\pm$ 0.012    &     0.695 $\pm$ 0.011     &     0.695 $\pm$ 0.008  \\
    push-reverse  & 0.640 $\pm$ 0.004          & 0.713 $\pm$ 0.023        &      0.717 $\pm$ 0.029    &     0.713 $\pm$ 0.023    &     0.738 $\pm$ 0.010     &     0.734 $\pm$ 0.010  \\
    push-p        & 0.696 $\pm$ 0.008          &      0.706 $\pm$ 0.003        &      0.699 $\pm$ 0.026    &     0.705 $\pm$ 0.021    &     0.714 $\pm$ 0.025     &     0.714 $\pm$ 0.022  \\
    
    \hline
   \end{tabular}
    }
\end{table*}

\section{Discussion}
In the first and second phases of RexNet, it learns users and items representations only based on the user-item matrix, however, in the suggested approach any method of representation learning can be used to infer in this phase. In this work we applied two different methods; Skip-gram and GloVe, and one important observation is the influence of these different approaches on the final performance. As we can observe by comparing tables \ref{result100k} and \ref{result1M}, using GloVe yields a better performance on Movielens 100k while Skip-gram results surpass those of GloVe in Movielens 1M. 
One reason of can be that the difference in sparsity of Movelens 100K and Movelens 1M. Table \ref{density_table} shows the density of both datasests. As we can see, Movielens 1M is sparser than Movielens 100K. The observed improvement in the performance of Glove in denser data set can be due to the fact that it heavily relies on global statistical information that it collects. In sparser datasets the available statistical information is not reliable enough and that may avoid Glove from inferring valid representations for some entities. But when we have a denser user-item matrix, GloVe's usage of statistical information helps it to infer embeddings that better represent the nature of entities. That can lead to better predictions in the next step and a higher overall recommendation performance.
Another observation is the significant superiority of RexNet over AutoRec. AutoRec uses an Auto Encoder to capture user's representations to predict user ratings, while RexNet captures item's and user's representations to learn their hierarchical features. To examine the ranking performance of AutoRec, we implemented it with the same setting of our algorithm to calculate NDCG. AutoRec's input is the sparse vector of users or items while RexNet first captures dense representations and then trains the neural network. It seems that having multiple layers in the neural network and also utilizing dense vectors of users and items helps RexNet to learn features that AutoRec is unable to capture. 

\begin{table*}[ht] \caption{Density of Movielens 100K and Movielens 1M.}
    \label{density_table}
    \centering
    \resizebox{\textwidth}{!}{

\begin{tabular}{ |c||c|c|c|c| }

    \hline
    DATASET &   USERS  & MOVIES & RATINGS &  DENSITY  \\
    \hline
    \hline
    MovieLens 100K & 943 & 1682 & 100,000 & 0.0630 \\
    \hline
    MovieLens 1M & 6040 & 3706 & 1,000,000 & 0.0446 \\
    \hline
   \end{tabular}
    }
\end{table*}

\section{Conclusion}
In this paper, we introduced a modular approach for collaborative filtering and our experiments showed that the suggested method outperforms the state-of-the-art CF algorithms in terms of NDCG. We showed that learning representations in a Collaborative Filtering environment can be done using well-known methods of word embedding such as Skip-gram and GloVe. The results showed that while Skip-gram performs better in sparser dataset while GloVe outperforms it in the denser one. We also argued that hierarchical feature extraction can be applied to recommendation systems and to do so, we introduced a deep parallel neural network architecture to capture hierarchical features of items and users in order to predict user preferences. Experimental results showed that the suggested approach performs quite well in two recommendation data sets. One future study can be validating different settings for the suggested structure to achieve better results. We tried GloVe and Skip-gram, that are well-known word embedding tools, for learning initial linear representations for users and items. The performance of the framework can possibly be further improved by designing special representation learning approaches which also takes collaborative filtering concerns into account.

\bibliography{rexnet}

\end{document}